\documentclass[aps,pra,twocolumn,superscriptaddress,showpacs]{revtex4}
\usepackage{graphicx}

\def\R{{\bf R}}
\def\C{{\bf C}}
\def\Z{{\bf Z}}

\def\im{{\rm i} }
\def\m{{\bf m}}

\begin{document}

\title{Wigner formula of rotation matrices and quantum walks}

\author{Takahiro Miyazaki}
\affiliation{Department of Physics,
Faculty of Science and Engineering,
Chuo University, Kasuga, Bunkyo-ku, Tokyo 112-8551, Japan}
\author{Makoto Katori} 
\email[]{katori@phys.chuo-u.ac.jp}
\affiliation{Department of Physics,
Faculty of Science and Engineering,
Chuo University, Kasuga, Bunkyo-ku, Tokyo 112-8551, Japan}
\author{Norio Konno}
\email[]{konno@ynu.ac.jp}
\affiliation{
Department of Applied Mathematics, 
Yokohama National University, 
79-5 Tokiwadai, Yokohama 240-8501, Japan}

\date{6 June 2007}

\begin{abstract}
Quantization of a random-walk model is performed by
giving a qudit (a multi-component wave function)
to a walker at site
and by introducing a quantum coin, which is 
a matrix representation of a unitary transformation.
In quantum walks, the qudit of walker is mixed 
according to the quantum coin at each time step,
when the walker hops to other sites.
As special cases of the quantum walks driven by 
high-dimensional quantum coins generally studied
by Brun, Carteret, and Ambainis, 
we study the models obtained by choosing rotation
as the unitary transformation, whose matrix
representations determine quantum coins.
We show that Wigner's $(2j+1)$-dimensional
unitary representations of rotations 
with half-integers $j$'s are useful to
analyze the probability laws of quantum walks.
For any value of half-integer $j$, 
convergence of all moments of walker's pseudovelocity
in the long-time limit is proved.
It is generally shown for the present models 
that, if $(2j+1)$ is even, the probability measure
of limit distribution
is given by a superposition of $(2j+1)/2$ terms
of scaled Konno's density functions,
and if $(2j+1)$ is odd, it is a superposition
of $j$ terms of scaled Konno's density functions
and a Dirac's delta function at the origin.
For the two-, three-, and four-component models,
the probability densities of 
limit distributions are explicitly
calculated and their dependence on the
parameters of quantum coins and on the initial
qudit of walker is completely determined.
Comparison with computer simulation results 
is also shown.

\end{abstract}

\pacs{03.65.-w,05.40.-a,03.67.-a}

\maketitle

\section{INTRODUCTION}

No one doubts the importance of random-walk models
in physics, mathematics and computer sciences.
In particular, when we explain basic concepts
of statistical physics \cite{Rei65},
stochastic processes in physics and chemistry \cite{vKam92},
and stochastic algorithms \cite{MR95},
introduction of random-walk models is very useful
and effective.
It is interesting to see that
systematic study of quantization of random walks
is not old \cite{ADZ93,Mey96,NV00,ABNVW01}.
As expected, the study of quantum walks is fruitful
and its results have been applied to solve
the transport problems in solid-state physics
of strongly correlated electron systems \cite{Oka05},
to derive non-Gaussian-type central limit theorems
in probability theory \cite{Kon02,Kon05a,GJS04},
and to invent new algorithms in the quantum information theory
\cite{Gro97,Amb04}.
The research field of quantum walks is growing
widely \cite{Kem03,TFMK03,Amb03,MBSS02,BCA03,Ken06} 
and mathematical understanding of the new models 
is becoming deeper \cite{Kon05b,GJS05,Kon06,Oba06,Kon07}
in recent years.

It should be noted here that, from the view-point of
standard quantum mechanics,
``to quantize random walks" is a contradictory concept,
since in quantum mechanics, time-evolution of 
a state vector $|\Psi(t) \rangle$, or 
a wave function $\Psi(x,t)$, is given by a
deterministic unitary transformation associated
with the Hamiltonian and probability concept
appears in the theory only when we perform observation
of physical quantities, {\it i.e.}, 
when we calculate the probability density
$p(x,t)=|\Psi(x,t)|^2$ at a given time.
On the other hand, random walk is a typical example of 
stochastic processes, 
in which we toss a coin to select a walk at each time step.
In an earlier paper \cite{KFK05}, it was shown that
the one-dimensional standard random walk can be
realized by a random-turn model \cite{Kon03}, in which
a coin is represented by a $2 \times 2$ stochastic matrix
and that, if we replace the matrix by a $2 \times 2$
unitary matrix, a one-dimensional quantum-walk model
is obtained.
This argument is not only heuristic but also
generic, since it implies that
quantization of random-walk models can be
done by introducing appropriate unitary matrices
such that they play the roles of ``quantum coins".
The obtained time-evolution of
quantum walk is described by 
{\it multi-component} version of quantum-mechanical
equation of motion.
For the standard quantum-walk model on
one-dimensional lattice 
$\Z=\{\cdots, -2, -1, 0, 1, 2, \cdots\}$ 
with the nearest-neighbor hopping, the equation is identified with the
Weyl equation for 
two-component wave functions \cite{KFK05}.
It should be remarked that such multi-component 
equations have been usually
used in {\it relativistic} quantum mechanics \cite{BES06}.

It should be noted that in order to discuss the relationship
between classical and quantum walks
Brun {\it et al.} showed a method to construct
discrete quantum-walk models for multi-component
wave functions driven by high-dimensional
quantum coins that are greater than $2 \times 2$ 
matrices \cite{BCA03}.
There they considered a $2^{M}$-dimensional space with
$M=2,3, \cdots$, whose basis is given by
tensor products of $M$ binary-vectors.
A quantum coin, which is first defined as an
abstract unitary transformation $\hat{U}$,
has a $2^{M} \times 2^{M}$ matrix
representation in the space \cite{BCA03}.
Their method is very general and useful
as well as the tensor-product method is
in the group representation theory
\cite{Geo99}.
See also Refs. \cite{MBSS02,BCA03b,VBBB05}
for tensor-product models.

In the present paper, we adopt the rotation
operator $\hat{R}$ specified by the Euler angles
$\alpha, \beta, \gamma$ 
(see Eq.(\ref{eqn:Rhat1}) in Sec.II) as the
unitary transformation $\hat{U}$
and introduce a family of quantum-walk models
on one-dimensional lattice $\Z$.
The two-dimensional representations of
$\hat{R}$ can be identified with $2 \times 2$
quantum-coin matrices used in the
standard quantum walks with qubits.
In the wave-number space ($k$-space)
the shift matrix $S(k)$ is generally
given by a matrix representation of
$\hat{R}$, if we set the parameters
$\alpha=-2k, \beta=\gamma=0$ (see Sec.III).
Our choice $\hat{U}=\hat{R}$ is thus
very suitable to study one-dimensional
quantum walk problems.

Instead of using the tensor-product representations
following Brun {\it et al.}, 
we will use in this paper the $(2j+1)$-dimensional
unitary representations
$R^{(j)}(\alpha, \beta, \gamma)$ of 
the rotation operator $\hat{R}$ with half-integers $j$'s
as quantum coins,
which are called the 
{\it rotation matrices} \cite{Mes75}.
The reason is the following.
The $2^{M}$-dimensional representations 
with $M \geq 2$ constructed by tensor products
are reducible;
By changing basis through appropriate
orthogonal transformations,
tensor-product matrices are block-diagonalized.
Wigner's theory showed that each block
with size $2j+1 \, (j=0, 1/2, 1, 3/2, \cdots)$
is given by a rotation matrix 
$R^{(j)}(\alpha, \beta, \gamma)$ \cite{Wig59}.
In other words, we will use the
irreducible representations of $\hat{R}$.
In Sec.IV, 
reducibility of the tensor-product models of quantum walks
to the present models will be generally 
shown and demonstrated using examples.

\begin{figure}[htpb]
\includegraphics[width=1\linewidth]{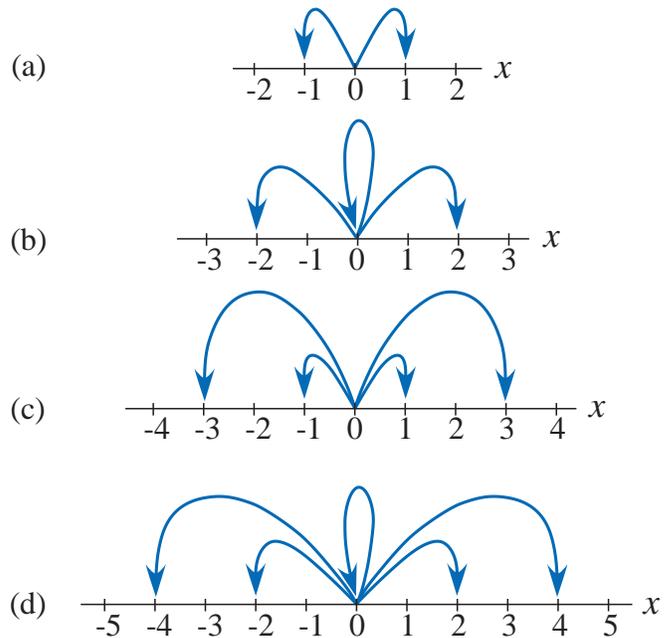}
\caption{(Color online)
Elementary hopping of quantum walker in the models
with (a) $j=1/2$ (two-component model),
(b) $j=1$ (three-component model), 
(c) $j=3/2$ (four-component model), and
(d) $j=2$ (five-component model).
When $(2j+1)$ is odd, the walker can stay at the same position
in a step, as shown by 
cyclic arrows at the origin in the cases (b) and (d).
\label{fig:mkkFig1}}
\end{figure}

In our model with $j$ and three parameters 
$(\alpha, \beta, \gamma)$,
a quantum walker is assumed to be at the origin at time $t=0$
with a $(2j+1)$-component qudit
\begin{equation}
\phi_{0}^{(j)}
=\left( \begin{array}{c}
q_j \cr q_{j-1} \cr \cdots \cr q_{-j+1} \cr q_{-j}
\end{array} \right)
\quad  \mbox{with} \quad
\sum_{m=-j}^{j} |q_{m}|^2=1,
\label{eqn:phi01}
\end{equation}
where $q_m \in \C$(complex numbers), $m=-j, -j+1, \cdots, j$.
At each time step $t=1,2, 3, \cdots$,
the components of qudit are mixed according to
a quantum coin $R^{(j)}(\alpha, \beta, \gamma)$
and the walker hops to $(2j+1)$ sites
on $\Z$, as illustrated by Fig.\ref{fig:mkkFig1}
for $j=1/2, 1, 3/2$ and 2
(see Eq.(\ref{eqn:Psix1}) in Sec.III).
When $j=1/2$, $R^{(1/2)}(\alpha, \beta, \gamma)$
can be identified with an element of SU(2) 
appropriately parameterized by the three variables
(the Cayley-Klein parameters) and the
model is reduced to the standard two-component
model \cite{KFK05}.
It should be noted here that, when $j$ is an integer
({\rm i.e.} $(2j+1)$ is odd),
the walker can stay at the same position in a step.

Using the method of \cite{GJS04}, we will prove
that any moments of pseudovelocity
of walker,
which is defined by $X_{t}/t$ 
(the position at time $t$, $X_{t}$, 
divided by $t$), converges in the long-time limit 
$t \to \infty$,
and show that the probability measure of 
limit distribution
is generally described by a superposition of
appropriately scaled forms of a function
\begin{equation}
\mu(x; a)=\frac{\sqrt{1-a^2}}
{\pi (1-x^2) \sqrt{a^2-x^2}}
{\bf 1}_{\{|x| < |a|\}},
\label{eqn:Konno1}
\end{equation}
where ${\bf 1}_{\{\omega\}}$ denotes the
indicator function of a condition $\omega$;
${\bf 1}_{\{\omega\}}=1$ if $\omega$ is satisfied
and ${\bf 1}_{\{\omega\}}=0$ otherwise,
and $a$ is a real parameter.
It is the density function
first introduced by Konno to describe the limit
distributions of the standard two-component quantum walks
in his weak limit-theorem \cite{Kon02,Kon05a}.
(As shown in Fig.\ref{fig:mkkFig2}(a) in Sec.VI,
$\mu(x;a)$ is inversed bell-shaped on a finite support
$x \in (-a,a)$ in big contrast with the Gaussian
distribution, which describes the diffusion scaling limit
of the classical random walks.)
More precisely speaking, when $(2j+1)$ is even, 
the probability density function of
limit distribution consists of
$(2j+1)/2$ terms of Konno's density functions (\ref{eqn:Konno1}),
and when $(2j+1)$ is odd, it does of
$j$ terms of Konno's and a point mass at the origin,
which corresponds to the positive probability
to retain the position of walker in a step
(see Eq.(\ref{eqn:nu1})).

The weight functions 
${\cal M}^{(j,m)}$ of these Konno's density functions
(and the weight $\Delta^{(j)}$ of 
Dirac's delta function at the origin,
when $(2j+1)$ is odd) in the superposition
depend on the parameters of quantum coin
and the $(2j+1)$ components of initial qudit (\ref{eqn:phi01})
of quantum walker.
We have found that the representation theory
of groups \cite{Geo99} is useful 
to calculate the weight functions. Especially
the Wigner formula for rotation matrices
\cite{Mes75,Wig59} is critical.
In this paper we give the explicit forms of weight functions
for $j=1/2, 1$ and 3/2 (two-, three- and four-component
models, respectively) and these results
imply that the weight functions 
${\cal M}^{(j,m)}$ are generally given by
polynomials.
Through these polynomials, the initial-qudit
dependence of the limit distribution of
pseudovelocity is completely determined.

This paper is organized as follows.
In Sec.II, Wigner formula of $(2j+1)$-dimensional irreducible
representation of the rotation group SO(3) is summarized.
The family of quantum-walk models associated with
the rotation matrices is defined
for quantum walkers with $(2j+1)$-component qudit
in Sec.III.
In Sec.IV we also introduce the tensor-product models
of one-dimensional quantum walks
associated with the rotation operator $\hat{R}$
following the general theory of Brun {\it et al.}
\cite{BCA03}.
There we show the reducibility of the tensor-product
models to our models.
Sec.V is devoted to proving the generalized weak limit-theorem
(convergence of all moments) for pseudovelocities
of quantum walks. There the polynomials,
which give the weights of Konno's density functions
and the point mass in the limit distributions,
are listed for $j=1/2, 1$ and 3/2, explicitly.
Comparison with computer simulation with
the present analytic results is given in Sec.VI.
In this last section, we also discuss relations between
our results and other multi-component 
models \cite{BCA03,BCA03b,IK05,IKS05,VBBB05,IKK04}
and possible future problems.

\section{WIGNER FORMULA OF $(2j+1)$-DIMENSIONAL
REPRESENTATIONS OF ROTATION GROUP}
Any rotation in the three-dimensional real space $\R^3$
is uniquely specified by three rotation angles
$\alpha, \beta, \gamma$ called the Euler angles.
In the quantum mechanics, the rotation with the Euler angles
$\alpha, \beta, \gamma$ is given by an operator of the
form (see, for instance, \cite{Mes75})
\begin{equation}
\hat{R}(\alpha, \beta, \gamma)=
e^{-\im \alpha \hat{J}_3} e^{-\im \beta \hat{J}_{2}}
e^{-\im \gamma \hat{J}_{3}},
\label{eqn:Rhat1}
\end{equation}
where $\hat{\bf J}
=(\hat{J}_{1}, \hat{J}_{2}, \hat{J}_{3})$ is the 
vector operator of angular momentum, whose elements satisfy the 
su(2) Lie algebra,
\begin{equation}
[\hat{J}_k, \hat{J}_{\ell}]
= \im \sum_{m=1}^{3} \varepsilon_{k \ell m}
\hat{J}_{m}, \quad
k, \ell=1,2,3,
\label{eqn:su2a1}
\end{equation}
with the completely antisymmetrical tensor 
with three indices
$\varepsilon_{k \ell m}$ \cite{Geo99}.
Let the ket-vector $|j, m \rangle, j=0,1/2,1,3/2, \cdots,
m=-j, -j+1, \cdots, j$, denote the normalized
eigenstates of $\hat{\bf J}^2=\sum_{k=1}^{3} \hat{J}_{k}^2$
and $\hat{J}_3$ such that
$\hat{\bf J}^2 |j, m \rangle=j(j+1) |j, m \rangle$,
and $\hat{J}_3 |j, m \rangle = m |j, m \rangle$.
(We set $\hbar=1$ in this paper.)
Then, for each fixed value of half-integer $j$, 
a $(2j+1) \times (2j+1)$ unitary 
matrix $R^{(j)}(\alpha, \beta, \gamma)=
(R^{(j)}_{m m'}(\alpha, \beta, \gamma))$ is defined 
with its elements
\begin{equation}
R^{(j)}_{m m'}(\alpha, \beta, \gamma)
=\langle j, m | \hat{R}(\alpha, \beta, \gamma)
|j, m' \rangle,
\label{eqn:Rj1}
\end{equation}
$m, m' =-j, -j+1, \cdots, j$.
We can show that
\begin{equation}
R^{(j)}_{m m'}(\alpha, \beta, \gamma)
=e^{-\im \alpha m} r^{(j)}_{m m'}(\beta)
e^{-\im \gamma m'}
\label{eqn:Rj2}
\end{equation}
with
\begin{eqnarray}
r^{(j)}_{m m'}(\beta)
&=& \sum_{\ell} \Gamma(j, m, m', \ell) \nonumber\\
&\times&
\left(\cos \frac{\beta}{2} \right)^{2j+m-m'-2 \ell}
\left(\sin \frac{\beta}{2} \right)^{2\ell+m'-m},
\label{eqn:rj1}
\end{eqnarray}
where
\begin{eqnarray}
&& \Gamma(j, m, m', \ell)\nonumber\\
&& =(-1)^{\ell}
\frac{\sqrt{(j+m)! (j-m)! (j+m')! (j-m')!}}
{(j-m'-\ell)! (j+m-\ell)! \ell ! (\ell +m'-m)!}.
\quad
\label{eqn:Gamma1}
\end{eqnarray}
In (\ref{eqn:rj1}) 
the summation $\sum_{\ell}$ extends over all integers
of $\ell$ for which the arguments of the factorials
are positive or null ($0!=1$).
The matrix (\ref{eqn:Rj2}) gives a
$(2j+1)$-dimensional irreducible representation of the
rotation group SO(3) and is called the rotation matrix.
Eq.(\ref{eqn:rj1}) is known as
the Wigner formula \cite{Mes75,Wig59}.
In the present paper, 
when we write matrices and vectors whose elements are
labeled by $m, m'$, we will assume that the indices $m$ and $m'$
run from $j$ to $-j$ in steps of $-1$.
In Appendix A, we give explicit expressions of matrices
$r^{(j)}(\beta)=
(r^{(j)}_{m m'}(\beta))$
for $j=1/2, 1$ and 3/2.

\section{QUANTUM-WALK MODELS WITH $(2j+1)$-COMPONENT QUDITS}

Here we propose a new family of models of
quantum walks on the one-dimensional lattice
$\Z$,
in which each walker has a $(2j+1)$-component qudit 
(\ref{eqn:phi01}).
In the previous paper \cite{KFK05},
we reported the weak limit-theorem for the two-component
model. That model is generated by a quantum coin represented by 
a matrix in SU(2)
\begin{eqnarray}
&& A=\left( \begin{array}{cc}
u e^{\im \theta} & \sqrt{1-u} e^{\im \phi} \cr
-\sqrt{1-u^2} e^{-\im \phi} & u e^{-\im \theta}
\end{array} \right),
\nonumber\\
&&
\qquad u \in [-1,1], \,
\theta, \phi \in [-\pi, \pi)
\label{eqn:A1}
\end{eqnarray}
and a spatial shift-operator on $\Z$, which is represented
by a matrix
\begin{equation}
S(k)=\left( \begin{array}{cc}
e^{\im k} & 0 \cr 0 & e^{-\im k} 
\end{array} \right), \quad
k \in [-\pi, \pi)
\label{eqn:Sk1}
\end{equation}
in the $k$-space.
If we compare these matrices with 
(\ref{eqn:Rj2}) with $j=1/2$ and (\ref{eqn:r1/2})
in Appendix A,
we find that they are the special cases of
$R^{(1/2)}(\alpha, \beta, \gamma)$;
\begin{eqnarray}
A &=& R^{(1/2)}(\pi-\theta-\phi, 2 {\rm arccos} (u),
-\pi - \theta+\phi),
\nonumber\\
S(k) &=& R^{(1/2)}(-2k, 0, 0).
\label{eqn:corresp1}
\end{eqnarray}
From the view point of the group theory,
we can give the following remark.
In \cite{KFK05} we used the fact that
SU(2)$\simeq {\bf S}^3$
($\equiv$ the three dimensional unit sphere in $\R^4$)
and the quantum coin $A$ was parameterized by three
real numbers, $u, \theta, \phi$
(the Cayley-Klein parameters), corresponding the
dimensionality 3 of the group space. 
On the other hand, 
we are now regarding the quantum coin $A$
as a two-dimensional representation of 
the rotation group SO(3), 
and thus the three parameters are identified with the
Euler angles for rotations in the three-dimensional
real-space $\R^3$.

This observation had led us to adopt 
the $(2j+1)$-dimensional representation of 
the rotation group, 
$R^{(j)}(\alpha, \beta, \gamma)$, as a quantum coin to
mix $(2j+1)$ components in qudit (\ref{eqn:phi01}).
The spatial shift-matrix is given by
$S^{(j)}(k)=R^{(j)}(-2k, 0, 0)
={\rm diag}(e^{2 \im j k}, e^{2 \im (j-1) k}, \cdots,
e^{-2 \im j k})$.

We assume at the initial time $t=0$ 
that the walker is located at the origin.
Then, in the $k$-space, the $(2j+1)$-component
wave function of the walker at time $t$ is given by
\begin{equation}
\hat{\Psi}^{(j)}(k,t)=\Big( V^{(j)}(k) \Big)^{t}
\phi_{0}^{(j)}, \quad
t=0,1,2, \cdots,
\label{eqn:Psih1}
\end{equation}
where
\begin{eqnarray}
V^{(j)}(k) &=& V^{(j)}(k; \alpha, \beta, \gamma)
\nonumber\\
&\equiv& S^{(j)}(k) R^{(j)}(\alpha, \beta, \gamma)
\nonumber\\
&=& R^{(j)}(\alpha-2k, \beta, \gamma).
\label{eqn:Vjk1}
\end{eqnarray}
The time evolution in the real space $\Z$ is
then obtained by performing the Fourier transformation,
\begin{eqnarray}
\Psi^{(j)}(x,t) &=& \int_{-\pi}^{\pi} \frac{dk}{2\pi}
\hat{\Psi}^{(j)}(k,t) e^{\im k x},\nonumber\\
\hat{\Psi}^{(j)}(k, t) &=&
\sum_{x \in \Z} \Psi^{(j)}(x,t) e^{-\im k x},
\label{eqn:Fourier1}
\end{eqnarray}
as
\begin{equation}
\Psi^{(j)}_{m}(x, t+1)
= \sum_{m'=-j}^{j} R^{(j)}_{m m'}
\Psi^{(j)}_{m'}(x+2m, t),
\, t=0,1,2, \cdots,
\label{eqn:Psix1}
\end{equation}
where $\Psi^{(j)}_{m}(x, t)$ denotes
the $m$-th component of the
$(2j+1)$-component wave function $\Psi^{(j)}(x,t)$.

Now the stochastic process of $(2j+1)$-component quantum walk
is defined on $\Z$ as follows.
Let $X^{(j)}_t$ be the position of the walker at time $t$.
The probability that we find the walker at site $x \in \Z$
at time $t$ is given by
\begin{equation}
{\rm Prob}(X^{(j)}_t=x)=
P(x,t)=
[\Psi^{(j)}(x,t)]^{\dagger} \Psi^{(j)}(x,t),
\label{eqn:Pxt1}
\end{equation}
where $[\Psi^{(j)}(x,t)]^{\dagger}$ is the
hermitian conjugate of $\Psi^{(j)}(x,t)$.
As shown in \cite{KFK05}, the $r$-th moment of $X^{(j)}_t$
is given by
\begin{eqnarray}
\langle (X^{(j)}_t)^{r} \rangle
&\equiv& \sum_{x \in \Z} x^r P(x,t) \nonumber\\
&=& \int_{-\pi}^{\pi} \frac{dk}{2\pi}
[\hat{\Psi}^{(j)}(k,t)]^{\dagger}
\left(\im \frac{d}{dk}\right)^{r}
\hat{\Psi}^{(j)}(k,t),
\nonumber\\
&& 
\qquad r=0,1,2, \cdots.
\label{eqn:moment1}
\end{eqnarray}

\section{TENSOR PRODUCT MODELS ASSOCIATED WITH
ROTATION OPERATOR AND THEIR REDUCIBILITY}
\subsection{Tensor product models}

In this section we use the notation
$|1 \rangle=|1/2, 1/2 \rangle$,
$|-1 \rangle =|1/2, -1/2 \rangle$
for the binary states with $j=1/2$.
Let $M \in \{2,3, 4, \cdots \}$.
For an $M$-component variable
$\m=(m_1, m_2, \cdots, m_M)$ with
$m_n \in \{-1, 1\}, 1 \leq n \leq M$,
we will write
$|\m|=\sum_{n=1}^{M} m_n$.

Following Brun {\it et al.} \cite{BCA03} we
consider the $2^M$-dimensional space
spanned by the bases 
$\{|\m \rangle \}_{\m \in \{-1,1\}^{M}}$,
which are defined as tensor products 
\begin{equation}
|\m \rangle=\prod_{n=1}^{M} |m_n \rangle.
\label{eqn:vecm1}
\end{equation}
Note that they are orthonormal;
$\langle \m | \m' \rangle
=\delta_{\m, \m'}
\equiv \prod_{n=1}^{M} \delta_{m_n m_n'}$.
Let
$$
{\rm R}^{[M]}_{\m, \m'}
(\alpha, \beta, \gamma)
=\langle \m | \hat{R}(\alpha, \beta, \gamma) | \m' \rangle,
$$
and define a $2^{M} \times 2^{M}$ matrix
${\rm R}^{[M]}(\alpha, \beta, \gamma)
=({\rm R}^{[M]}_{\m, \m'}(\alpha, \beta, \gamma) )
_{\m, \m' \in \{-1, 1\}^{M}}$.
By definition of tensor products \cite{Geo99},
\begin{eqnarray}
{\rm R}^{[M]}_{\m, \m'}
(\alpha, \beta, \gamma)
&=& \prod_{n=1}^{M} R^{(1/2)}_{m_n/2, m_n'/2}
(\alpha, \beta, \gamma)
\nonumber\\
&=& e^{-\im \alpha |\m|/2 -\im \gamma |\m'|/2}
\prod_{n=1}^{M} r^{(1/2)}_{m_n/2, m_n'/2}(\beta),
\nonumber
\end{eqnarray}
where $r^{(1/2)}(\beta)=
(r^{(1/2)}_{m/2, m'/2})_{m, m' \in \{-1,1\}}$
is given by (\ref{eqn:r1/2}) 
in Appendix A.
This gives a $2^{M}$-dimensional 
tensor-product representation of the
rotation group.
Define the shift matrix in the $k$-space by
$
{\rm S}^{[M]}_{\m, \m'}(k)
={\rm R}^{[M]}_{\m, \m'}(-2k, 0, 0)
=e^{\im k |\m|} \delta_{\m \m'},
$
and set
\begin{eqnarray}
{\rm V}^{[M]}(k)
&=& {\rm S}^{[M]}(k) {\rm R}^{[M]}(\alpha, \beta, \gamma)
\nonumber\\
&=& {\rm R}^{[M]}(\alpha-2k, \beta, \gamma).
\label{eqn:rmV1}
\end{eqnarray}
Following the general theory of
Brun {\it et al.} \cite{BCA03},
the wave function in the $k$-space at time $t$
of the one-dimensional quantum walk
associated with the above tensor-product representation
of SO(3) will be given by
\begin{equation}
\hat{\Phi}^{[M]}(k, t)
= \Big( {\rm V}^{[M]}(k) \Big)^{t}
\varphi_{0}^{[M]}, \quad t=0,1,2, \cdots,
\label{eqn:Phi1}
\end{equation}
where the initial state is given by the
$2M$-component qudit
\begin{equation}
\varphi_0^{[M]} = \bigotimes_{n=1}^{M} 
\left( \begin{array}{c}
Q_n^{+} \cr Q_n^{-} 
\end{array} \right), \quad
Q_n^{+}, Q_n^{-} \in \C
\label{eqn:qudit1}
\end{equation}
with the normalization
$\sum_{n=1}^{M}(|Q_n^{+}|^2+|Q_n^{-}|^2)=1$.
The real-space wave function is then given by its Fourier 
transformation
\begin{equation}
\Phi^{[M]}(x,t)=\int_{-\pi}^{\pi} \frac{dk}{2 \pi}
\hat{\Phi}^{[M]}(k,t) e^{\im k x}.
\label{eqn:real1}
\end{equation}

Let $Y_t^{[M]}, t=0,1,2, \cdots,$ be the position
of the walker at time $t$ of this tensor-product model.
The probability distribution function is defined by
\begin{equation}
P^{[M]}(x,t) \equiv
{\rm Prob}(Y^{[M]}_t=x)
= [\Phi^{[M]}(x,t)]^{\dagger} \Phi^{[M]}(x,t).
\label{eqn:probM1}
\end{equation}
The initial position of this quantum walk is the origin,
$Y_0^{[M]}=0$, and the walker has the initial qudit
(\ref{eqn:qudit1}).

\subsection{Reduction of quantum-walk models}

Irreducible representations of the rotation group
are given in the spaces spanned by
$|j, m \rangle$, $j=0,1/2,1,3/2, \cdots$,
$m=-j, -j+1, \cdots, j-1, j$
\cite{Wig59,Geo99}.
The two kinds of bases 
$\{|\m \rangle \}$ and $\{|j, m \rangle^{\ell_j} \}$
are related through
\begin{equation}
|\m \rangle
= \sum_{j} \sum_{\ell_{j}=1}^{d_{j}} \sum_{m_{j}}
|j, m_j \rangle^{\ell_j}
{\rm K}_{(j, m_j)^{\ell_j}, \m}
\label{eqn:transf1}
\end{equation}
with
\begin{equation}
{\rm K}^{[M]}_{(j, m_j)^{\ell_j}, \m}
=^{\ell_j}\!\langle j, m_j | \m \rangle,
\label{eqn:matrixK1}
\end{equation}
where $d_{j}$ is the multiplicity of the 
$(2j+1)$-dimensional irreducible representations included in 
the $2^M$-dimensional tensor-product representation
and for each $j$
an index $\ell_j$ runs from 1 to $d_j$.
Remark that
$\sum_j d_j (2j+1)=2^{M}$.
Define the $2^{M}$-dimensional matrix
$ {\rm K}^{[M]}=({\rm K}^{[M]}_{(j, m_j)^{\ell_j}, \m})$,
which is an orthogonal matrix;
$({\rm K}^{[M]})^{-1}=^{t}\!{\rm K}^{[M]}$.

Then we see
\begin{eqnarray}
&&{\rm K}^{[M]} {\rm R}^{[M]}(\alpha-2k, \beta, \gamma)
({\rm K}^{[M]})^{-1} \nonumber\\
&=& \bigoplus_{j, \ell_j}
R^{(j, \ell_j)}(\alpha-2k, \beta, \gamma)
\nonumber\\
&=& \bigoplus_j
\Big\{ R^{(j,1)}(\alpha-2k, \beta, \gamma)
\oplus \cdots \nonumber\\
&& \qquad \qquad 
\oplus R^{(j, d_{j})}(\alpha-2k, \beta, \gamma) \Big\},
\label{eqn:directsum}
\end{eqnarray}
where $R^{(j, \ell_{j})}, \ell=1,2, \cdots, d_{j}$,
are $d_{j}$ copies of the $(2j+1)$-dimensional
irreducible representation of rotation group 
explained in Sec.II.
That is, 
${\rm R}^{[M]}(\alpha-2k, \beta, \gamma)$
can be block-diagonalized into
a direct sum of rotation matrices. 
Note that direct sums in (\ref{eqn:directsum})
and equations below are taken only over
$j$'s such that 
${\rm K}^{[M]}_{(j, m_j)^{\ell_j}, \m} 
\not=0$. 
(See the examples 
in the following subsection.)
For (\ref{eqn:rmV1}) it implies that
\begin{equation}
({\rm V}^{[M]}(k))^t
= ({\rm K}^{[M]})^{-1} \bigoplus_{j, \ell_j} (V^{(j, \ell_j)}(k))^{t}
{\rm K}^{[M]},
\label{eqn:dsum1}
\end{equation}
where $V^{(j, \ell_j)}(k)$ is the $\ell_j$-th
copy of (\ref{eqn:Vjk1}).

Let
\begin{equation}
{\rm K}^{[M]} \varphi_0^{[M]}
= \bigoplus_{j, \ell_j}
\varphi_{0}^{(j, \ell_{j})},
\label{eqn:initial0}
\end{equation}
and define
\begin{equation}
p^{(j, \ell_j)}=[\varphi_0^{(j, \ell_j)}]^{\dagger}
\varphi_0^{(j, \ell_j)}.
\label{eqn:pjl1}
\end{equation}
Then it is easy to prove that
the probability distribution function
(\ref{eqn:probM1}) is decomposed as
\begin{equation}
P^{[M]}(x,t)=
\sum_{j} \sum_{\ell_j=1}^{d_j}
p^{(j, \ell_j)}
P^{(j, \ell_j)}(x, t),
\label{eqnn:PMdecom1}
\end{equation}
where $P^{(j, \ell_{j})}(x,t)$
is the probability distribution function
of our quantum-walk model
introduced in Sec.III, whose initial
$(2j+1)$-component qudit is given by
\begin{equation}
\phi_0^{(j, \ell_j)}
= \frac{1}{\sqrt{p^{(j, \ell_j)}}}
\varphi_0^{(j, \ell_j)}.
\label{eqn:phi0jl}
\end{equation}
By this formula,
the probability laws of
quantum walks of the tensor-product models
are completely determined by
those of the models studied in the present paper.

\subsection{Examples}
\noindent \underline{$M=2$ case} \quad
We set the $2^2=4$ states 
$\{ |(m_1, m_2) \rangle \}_{(m_1, m_2) \in \{-1,1 \}^2}$
in the order
$|(1,1) \rangle$, $|(1,-1) \rangle$, $|(-1,1) \rangle$,
$|(-1,-1) \rangle$.
Then we have
$
{\rm R}^{[2]}_{\m, \m'}(\alpha, \beta, \gamma)
= e^{-\im \alpha(m_1+m_2) - \im \gamma(m_1'+m_2')}
{\rm r}^{[2]}_{\m, \m'}
$
with
\begin{equation}
{\rm r}^{[2]} = \Big({\rm r}^{[2]}_{\m, \m'} \Big)
= \left( \begin{array}{cccc}
c^2 & -cs & -cs & s^2 \cr
cs & c^2 & -s^2 & -cs \cr
cs & -s^2 & c^2 & -cs \cr
s^2 & cs & cs & c^2
\end{array} \right).
\label{eqn:r_2}
\end{equation}
The shift matrix is given in the $k$-space by
\begin{equation}
{\rm S}^{[2]}(k) = {\rm R}^{[2]}(-2k, 0, 0)
= \left( \begin{array}{cccc}
e^{2 \im k} & 0 & 0 & 0 \cr
0 & 1 & 0 & 0 \cr
0 & 0 & 1 & 0 \cr
0 & 0 & 0 & e^{-2 \im k} 
\end{array} \right).
\label{eqn:S2}
\end{equation}
From the above four states, 
$\{|j,m \rangle : j=1,0 \}$ are
obtained by the highest weight construction
(see, for example, \cite{Geo99})
and we find the transformation matrix ${\rm K}^{[2]}$ as
\begin{equation}
{\rm K}^{[2]} 
= \Big( \langle j, m | (m_1, m_2) \rangle \Big)
=\left( \begin{array}{cccc}
1 & 0 & 0 & 0 \cr
0 & \sqrt{1/2} & \sqrt{1/2} & 0 \cr
0 & 0 & 0 & 1 \cr
0 & \sqrt{1/2} & -\sqrt{1/2} & 0
\end{array} \right).
\label{eqn:K22}
\end{equation}
It is easy to confirm that
\begin{eqnarray}
{\rm K}^{[2]} {\rm r}^{[2]}(\beta)
({\rm K}^{[2]})^{-1}
&=& \left( \begin{array}{cccc}
c^2 & -\sqrt{2}cs & s^2 & 0 \cr
\sqrt{2} cs & 2c^2-1 & - \sqrt{2} cs & 0 \cr
s^2 & \sqrt{2} cs & c^2 & 0 \cr
0 & 0 & 0 & 1 
\end{array} \right) \nonumber\\
&=& r^{(1)}(\beta) \oplus 1,
\label{eqn:case2b}
\end{eqnarray}
where $r^{(1)}(\beta)$ is given by (\ref{eqn:r1/2})
in Appendix A.
This implies 
$
{\rm K}^{[2]}
{\rm V}^{[2]}(k) ({\rm K}^{[2]})^{-1}
= V^{(1)}(k) \oplus 1.
$
This decomposition will be
symbolically denoted as
\begin{equation}
2 \otimes 2 = 3 \oplus 1.
\label{eqn:case2d}
\end{equation}

\begin{widetext}
\noindent\underline{$M=3$ case} \quad
We set the $2^3=8$ states 
$\{|(m_1, m_2, m_3) \rangle\}_{
(m_1, m_2, m_3) \in \{-1, 1\}^{3}}$
in the order
$|(1,1,1) \rangle$, $|(1,1,-1) \rangle$, $|(1,-1,1) \rangle$, 
$|(1,-1,-1) \rangle$, $|(-1,1,1) \rangle$, $|(-1,1,-1) \rangle$,
$|(-1,-1,1) \rangle$, $|(-1,-1,-1) \rangle$.
Then we have
$
{\rm R}^{[3]}_{\m, \m'}(\alpha, \beta. \gamma)
= e^{-\im \alpha(m_1+m_2+m_3) - \im \gamma(m_1'+m_2'+m_3')}
{\rm r}^{[3]}_{\m, \m'}
$
with
\begin{equation}
{\rm r}^{[3]} = \Big({\rm r}^{[3]}_{\m, \m'} \Big)
= \left( \begin{array}{cccccccc}
c^3 & -c^2 s & -c^2 s & c s^2 & -c^2 s & c s^2 & c s^2 & -s^3 \cr
c^2 s & c^3 & -c s^2 & -c^2 s & -c s^2 & -c^2 s & s^3 & cs^2 \cr
c^2 s & -cs^2 & c^3 & -c^2 s & -cs^2 & s^3 & -c^2s & cs^2 \cr
cs^2 & c^2 s & c^2 s & c^3 & -s^3 & -cs^2 & -cs^2 & -c^2 s \cr
c^2 s & -cs^2 & -cs^2 & s^3 & c^3 & -c^2s & -c^2s & cs^2 \cr
cs^2 & c^2 s & -s^3 & -cs^2 & c^2 s & c^3 & -cs^2 & -c^2s \cr
cs^2 & -s^3 & c^2 s & -cs^2 & c^2 s & -cs^2& c^3 & -c^2 s \cr
s^3 & cs^2 & cs^2 & c^2 s & c s^2 & c^2 s & c^2 s & c^3
\end{array} \right).
\label{eqn:r_3}
\end{equation}
The shift matrix is given in the $k$-space as
\begin{equation}
{\rm S}^{[3]}(k) = {\rm R}^{[3]}(-2k, 0, 0)
= \left( \begin{array}{cccccccc}
e^{3 \im k} & & & & & & & \cr
& e^{\im k} & & & & & & \cr
& & e^{\im k} & & & & & \cr
& & & e^{-\im k} & & & & \cr
& & & & e^{\im k} & & & \cr
& & & & & e^{-\im k} & & \cr
& & & & & & e^{-\im k} & \cr
& & & & & & & e^{-3 \im k}
\end{array} \right).
\label{eqn:S3}
\end{equation}
By the highest weight construction, we find 
in this case that we obtain a pair of
two-dimensional subspaces in addition to one 
four-dimensional subspace in the decomposition;
\begin{equation}
2 \otimes 2 \otimes 2 = 4 \oplus 2 \oplus 2.
\label{eqn:decomp2}
\end{equation}
That is, the multiplicities are
$$
d_{3/2}=1 \quad \mbox{and} \quad
d_{1/2}=2.
$$
The orthogonal matrix is
determined as
\begin{equation}
{\rm K}^{[3]}
=\left( \begin{array}{cccccccc}
1 & 0 & 0 & 0 & 0 & 0 & 0 & 0 \cr
0 & \sqrt{1/3} & \sqrt{1/3} & 0 & \sqrt{1/3} & 0 & 0 & 0 \cr
0 & 0 & 0 & \sqrt{1/3} & 0 & \sqrt{1/3} & \sqrt{1/3} & 0 \cr
0 & 0 & 0 & 0 & 0 & 0 & 0 & 1 \cr
0 & \sqrt{1/6} & - \sqrt{2/3} & 0 & \sqrt{1/6} & 0 & 0 & 0 \cr
0 & 0 & 0 & -\sqrt{1/6} & 0 & \sqrt{2/3} & - \sqrt{1/6} & 0 \cr
0 & -\sqrt{1/2} & 0 & 0 & \sqrt{1/2} & 0 & 0 & 0 \cr
0 & 0 & 0 & -\sqrt{1/2} & 0 & 0 & \sqrt{1/2} & 0 
\end{array} \right).
\label{eqn:K32}
\end{equation}
We can see then
\begin{eqnarray}
{\rm K}^{[3]} {\rm r}^{[3]}(\beta)
({\rm K}^{[3]})^{-1}
&=& \left( \begin{array}{cccccccc}
c^3 & -\sqrt{3} c^2 s & \sqrt{3} cs^2 & - s^3 & 0 & 0 & 0 & 0 \cr
\sqrt{3} c^2 s & c(c^2-2s^2) & -s(2c^2-s^2) & \sqrt{3} cs^2 & 
0 & 0 & 0 & 0 \cr
\sqrt{3} c s^2 & s(2c^2-s^2) & c(c^2-2s^2) & -\sqrt{3} c^2 s &
0 & 0 & 0 & 0 \cr
s^3 & \sqrt{3} cs^2 & \sqrt{3} c^2s & c^3 &
0 & 0 & 0 & 0 \cr
0 & 0 & 0 & 0 & c & -s & 0 & 0 \cr
0 & 0 & 0 & 0 & s & c & 0 & 0 \cr
0 & 0 & 0 & 0 & 0 & 0 & c & -s \cr
0 & 0 & 0 & 0 & 0 & 0 & s & c
\end{array} \right) \nonumber\\
&=& r^{(3/2)}(\beta)
\oplus r^{(1/2, 1)}(\beta)
\oplus r^{(1/2, 2)}(\beta),
\label{eqn:case3b}
\end{eqnarray}
where $r^{(3/2)}(\beta)$ is (\ref{eqn:r3/2})
and both of $r^{(1/2,1)}(\beta)$ and
$r^{(1/2,2)}(\beta)$ are identified with (\ref{eqn:r1/2})
in Appendix A.
This implies
\begin{equation}
{\rm K}^{[3]}
{\rm V}^{[3]}(k) ({\rm K}^{[3]})^{-1}
= V^{(3/2)}(k) \oplus V^{(1/2,1)}(k) \oplus V^{(1/2, 2)}(k).
\label{eqn:case3c}
\end{equation}
\end{widetext}
\section{LIMIT DISTRIBUTIONS OF QUANTUM WALKERS}
\subsection{Decomposition of time-evolution matrix}
A key lemma for the following analysis of the
quantum-walk models is the fact that
the time-evolution matrix $V^{(j)}(k)$ defined by 
(\ref{eqn:Vjk1}) is decomposed into the three rotation matrices
$R^{(j)}$'s of the form
\begin{eqnarray}
V^{(j)}(k) &=& R^{(j)}(\phi(k), \theta(k), 0)
R^{(j)}(-p(k), 0, 0) \nonumber\\
&& \qquad \times
[R^{(j)}(\phi(k), \theta(k), 0)]^{\dagger},
\label{eqn:decom1}
\end{eqnarray}
where $\phi(k), \theta(k)$ and $p(k)$ are
related with the Euler angles $\alpha, \beta, \gamma$
and the wave number $k$ by 
\begin{eqnarray}
\frac{1}{2} \{ (\alpha-2k)-\gamma \}
&=& \phi(k)+\frac{\pi}{2}, \nonumber\\
\tan \frac{1}{2} \{(\alpha-2k)+\gamma\}
&=& -\tan \frac{p(k)}{2} \cos \theta(k), \nonumber\\
\sin \frac{\beta}{2} &=& \sin \frac{p(k)}{2} \sin \theta(k).
\label{eqn:decom2}
\end{eqnarray}
We give the proof of this formula in Appendix B.

The formula (\ref{eqn:decom1}) means that the time-evolution 
matrix $V^{(j)}(k)$ can be diagonalized to
$R^{(j)}(-p(k),0,0)$ by a unitary transformation
given by $R^{(j)}(\phi(k), \theta(k), 0)$.
Indeed $R^{(j)}(-p(k), 0,0)$ is a diagonal matrix
$R^{(j)}(-p(k), 0, 0)
={\rm diag}(e^{\im j p(k)}, e^{\im (j-1) p(k)}, \cdots,
e^{-\im j p(k)})$
and, by the unitarity of $R^{(j)}$, 
$[R^{(j)}]^{\dagger}=[R^{(j)}]^{-1}$,
(\ref{eqn:Psih1}) is written as
\begin{eqnarray}
\hat{\Psi}^{(j)}(k, t) &=&
R^{(j)}(\phi(k), \theta(k), 0)
\nonumber\\
&\times&
\left( \begin{array}{cccc}
e^{\im t j p(k)} & & & \cr
 & e^{\im t (j-1) p(k)} & & \cr
 & & \ddots & \cr
 & & & e^{-\im t j p(k)} 
 \end{array} \right) \nonumber\\
 && \qquad \times
[R^{(j)}(\phi(k), \theta(k), 0)]^{\dagger}
\phi_{0}^{(j)}
\nonumber\\
&=& \sum_{m=-j}^{j} e^{\im t m p(k)}
{\bf v}^{(j)}_{m}(k) C_{m}^{(j)}(k),
\label{eqn:Psih2}
\end{eqnarray}
where ${\bf v}^{(j)}_{m}(k)$ is the $m$-th column
vector in the matrix $R^{(j)}(\phi(k), \theta(k), 0)$,
$$
 {\bf v}^{(j)}_{m}(k)
 =\left( \begin{array}{c}
 R^{(j)}_{j m}(\phi(k), \theta(k), 0) \cr
 R^{(j)}_{j-1 \, m}(\phi(k), \theta(k), 0) \cr
 \cdots \cr
 R^{(j)}_{-j m}(\phi(k), \theta(k), 0) 
 \end{array} \right)
$$
and
\begin{equation}
C^{(j)}_{m}(k) \equiv [{\bf v}^{(j)}_{m}(k)]^{\dagger}
\phi^{(j)}_0 = \sum_{m'=-j}^{j} 
\overline{R^{(j)}_{m' m}}
(\phi(k), \theta(k), 0) q_{m'},
\label{eqn:Cj1}
\end{equation}
where $\overline{z}$ denotes the complex conjugate
of a complex number $z$.

The expansion (\ref{eqn:Psih2}) gives
\begin{eqnarray}
&& \left(\im \frac{d}{dk} \right)^{r}
\hat{\Psi}^{(j)}(k,t)
\nonumber\\
&& \quad = \sum_{m=-j}^{j} \left(-m
\frac{dp(k)}{dk} \right)^{r} 
e^{\im t m p(k)}
{\bf v}^{(j)}_m(k) C_{m}^{(j)}(k) \, t^{r}
\nonumber\\
&& \qquad \qquad 
+{\cal O}(t^{r-1}).
\nonumber
\end{eqnarray}
Since $R^{(j)}$ is unitary, its column vectors 
make a set of orthonormal vectors,
$
[{\bf v}^{(j)}_m(k)]^{\dagger} {\bf v}^{(j)}_{m'}(k)
=\delta_{m m'}.
$
Then we have
\begin{eqnarray}
&& [\hat{\Psi}^{(j)}(k,t)]^{\dagger}
\left(\im \frac{d}{dk}\right)^{r}
\hat{\Psi}^{(j)}(k,t) \nonumber\\
&=& \sum_{m=-j}^{j} 
\left(- m \frac{dp(k)}{dk} \right)^{r}
|C^{(j)}_{m}(k)|^2 t^r
+{\cal O}(t^{r-1}),
\nonumber
\end{eqnarray}
and thus (\ref{eqn:moment1}) gives
the following expression for
moments of pseudovelocity 
$X^{(j)}_{t}/t$ in the long-time limit \cite{GJS04,KFK05},
\begin{eqnarray}
&& \lim_{t \to \infty}
\Big\langle \Big(\frac{X^{(j)}_{t}}{t} \Big)^{r} \Big\rangle
\nonumber\\
&=& \sum_{m: 0 < m \leq j}
\int_{-\pi}^{\pi} \frac{dk}{2\pi}
\Big\{ (-1)^{r}|C^{(j)}_{m}(k)|^2
+|C^{(j)}_{-m}(k)|^2 \Big\} \nonumber\\
&& \qquad \qquad \qquad \times
\left( m \frac{dp(k)}{dk} \right)^{r},
\label{eqn:limit1}
\end{eqnarray}
$r=1,2,3, \cdots$, 
where the summation is taken over 
$m=1/2, 3/2, \cdots, j$, if $j$ is a half of odd number,
and $m=1,2, \cdots, j$, if $j$ is a positive integer.
Here it should be noted that, when $j$ is a positive
integer, $m=0$ mode exists, but it does not contribute
to any moment of order $r=1,2,3, \cdots$ in 
(\ref{eqn:limit1}).
The $m=0$ mode comes from the fact that the walker can stay
at the same position in a step, when
$(2j+1)$ is odd, and its contribution to the limit
distribution will be described by a point mass
at the origin (see Sec.V.C).

\subsection{Planar orbits in parameter space and integrals}

The equations (\ref{eqn:decom2}) define a one-parameter
family (with parameter $k$) of transformations
from the Euler angles $(\alpha, \beta, \gamma)$ 
to $(p, \theta, \phi)$.
More explicitly, we can find the following equations
from (\ref{eqn:decom2}) (see Appendix B),
\begin{eqnarray}
\label{eqn:mapA1}
\cos \frac{p(k)}{2} &=& \cos \frac{\beta}{2}
\cos \frac{1}{2}(\alpha+\gamma-2k), \\
\sin \frac{p(k)}{2} &=& \sqrt{
1-\cos^2(\beta/2) \cos^2\{(\alpha+\gamma-2k)/2\}}, \nonumber\\
\label{eqn:mapA2}
\\
\cos \theta(k) &=& -\frac{\cos(\beta/2) \sin\{(\alpha+\gamma-2k)/2\}}
{\sqrt{1-\cos^2(\beta/2) \cos^2\{(\alpha+\gamma-2k)/2\}}}, 
\nonumber\\
\label{eqn:mapA3}
\\
\sin \theta(k) &=& \frac{\sin(\beta/2)}
{\sqrt{1-\cos^2(\beta/2) \cos^2\{(\alpha+\gamma-2k)/2\}}}, 
\nonumber\\
\label{eqn:mapA4}
\\
\label{eqn:mapA5}
\phi(k) &=& \frac{1}{2}(\alpha-\gamma-2k-\pi).
\end{eqnarray}
Following the argument given in \cite{KFK05},
we consider a vector ${\bf p}(k)=(p_{1}(k), p_{2}(k), p_{3}(k))$
in the three-dimensional parameter space defined by
\begin{eqnarray}
p_{1}(k) &=& p(k) \sin \theta(k) \cos \phi(k), \nonumber\\
p_{2}(k) &=& p(k) \sin \theta(k) \sin \phi(k), \nonumber\\
p_{3}(k) &=& p(k) \cos \theta(k).
\label{eqn:vecp1}
\end{eqnarray}
Let
\begin{eqnarray}
\hat{\bf e}_{1} &=& (-\sin \gamma, -\cos \gamma, 0), \nonumber\\
\hat{\bf e}_{2} &=& (\sin \frac{\beta}{2} \cos \gamma, 
-\sin\frac{\beta}{2} \sin \gamma, - \cos\frac{\beta}{2}), \nonumber\\
\hat{\bf e}_{3} &=& (\cos\frac{\beta}{2} \cos \gamma,
-\cos \frac{\beta}{2} \sin \gamma, \sin\frac{\beta}{2}).
\label{eqn:e1e2e3}
\end{eqnarray}
Using (\ref{eqn:mapA1})-(\ref{eqn:mapA5}), it is easy to confirm
the fact that
$$
{\bf p}(k) \perp \hat{\bf e}_3
\quad \mbox{for all} \quad k \in [-\pi, \pi),
$$
which implies that ${\bf p}(k)$ draws an
orbit on a plane including the origin,
whose normal vector is $\hat{\bf e}_{3}$ in the parameter space.
On this orbital plane, we define the angle $\chi$
by 
$\cos \chi=\hat{\bf p}(k) \cdot \hat{\bf e}_{1}$,
where $\hat{\bf p}(k)={\bf p}(k)/p(k)$.
Then we have the relations
\begin{eqnarray}
\label{eqn:coschi1}
\cos \chi &=& \frac{\sin(\beta/2) \cos\{(\alpha+\gamma-2k)/2\}}
{\sqrt{1-\cos^2(\beta/2) \cos^2\{(\alpha+\gamma-2k)/2\}}},
\qquad \\
\label{eqn:sinchi1}
\sin \chi &=& \frac{\sin\{(\alpha+\gamma-2k)/2\}}
{\sqrt{1-\cos^2(\beta/2) \cos^2\{(\alpha+\gamma-2k)/2\}}}.
\qquad 
\end{eqnarray}
Comparing (\ref{eqn:coschi1}) with (\ref{eqn:mapA1}) and 
(\ref{eqn:mapA2}), the equation of the orbit
on the plane is determined of essentially the same form
as reported in \cite{KFK05},
\begin{equation}
\tan \frac{p(k)}{2}
=\tan \frac{\beta}{2} \frac{1}{\cos \chi}.
\label{eqn:orbit1}
\end{equation}

As pointed by \cite{KFK05},
the integral with respect to the wave number $k$
in (\ref{eqn:limit1}) is mapped to the
curvilinear integration along the orbit 
with respect to the angle $\chi$ through the relations
(\ref{eqn:coschi1}) and (\ref{eqn:sinchi1}), or their
inverted forms
\begin{eqnarray}
\label{eqn:coschi2}
\cos \frac{1}{2}(\alpha+\gamma-2k)
&=& \frac{\cos \chi}{\sqrt{1-\cos^2(\beta/2) \sin^2 \chi}}, \\
\label{eqn:sinchi2}
\sin \frac{1}{2}(\alpha+\gamma-2k)
&=& \frac{\sin \chi \sin(\beta/2)}
{\sqrt{1-\cos^2(\beta/2) \sin^2 \chi}}.
\end{eqnarray}
The Jacobian associated with the map
$k \mapsto \chi$ is obtained as
\begin{equation}
J \equiv \left| \frac{dk}{d \chi} \right|
=\frac{\sin(\beta/2)}{1-\cos^2(\beta/2) \sin^2 \chi}.
\label{eqn:Jacobian1}
\end{equation}
From (\ref{eqn:mapA2}), we have
$$
p(k)=2 \, {\rm arccos}
\left\{ \cos \frac{\beta}{2} 
\cos \frac{1}{2}(\alpha+\gamma-2k) \right\},
$$
and then
\begin{equation}
\frac{dp(k)}{dk}= - 2 \cos \frac{\beta}{2} \sin \chi,
\label{eqn:dpdk1}
\end{equation}
where the formula
$(d/dx) {\rm arccos} x=\mp 1/\sqrt{1-x^2}$ 
has been used.
The long-time limit (\ref{eqn:limit1})
of moments of pseudovelocity is now expressed as
\begin{equation}
\lim_{t \to \infty}
\Big\langle \Big( \frac{X^{(j)}_t}{t} \Big)^{r} 
\Big\rangle
= \sum_{m: 0 < m \leq j} I^{(j)}_{m}(r),
\quad r=1,2,3, \cdots,
\label{eqn:limit2}
\end{equation}
where
\begin{eqnarray}
I^{(j)}_{m}(r)
&=& \int_{-\pi}^{\pi} \frac{d \chi}{2\pi}
\frac{\sin(\beta/2)}{1-\cos^2(\beta/2) \sin^2 \chi}
\nonumber\\
&& \times
\Big\{|\hat{C}^{(j)}_{m}(\chi)|^2
+(-1)^{r}|\hat{C}^{(j)}_{-m}(\chi)|^2 \Big\}
\nonumber\\
&& \quad \times 
\left( 2 m \cos \frac{\beta}{2} \sin \chi \right)^{r}
\label{eqn:limit3}
\end{eqnarray}
with
$\hat{C}^{(j)}_{\pm m}(\chi) \equiv
C^{(j)}_{\pm m}(k(\chi))$.

\subsection{Superposition of scaled Konno's density functions}

In each integral $I^{(j)}_{m}(r)$,
we change the variable of integral from $\chi$ to $y$ by
\begin{equation}
y=2 m \cos \frac{\beta}{2} \sin \chi.
\label{eqn:y1}
\end{equation}
If we assume that by this change of variable
$\hat{C}^{(j)}_m(\chi)$ is replaced by $c^{(j)}_{m}(y)$,
the integral is written as
\begin{eqnarray}
I^{(j)}_{m}(r)
&=& \frac{1}{2m} \int_{-\infty}^{\infty} dy \, y^{r}
\mu \left( \frac{y}{2m}; \cos \frac{\beta}{2} \right)
\nonumber\\
&& \times
\Big\{|c^{(j)}_{m}(y)|^2
+(-1)^{r}|c^{(j)}_{-m}(y)|^2 \Big\}.
\label{eqn:Ijm1}
\end{eqnarray}
Here $\mu(x; a)$ is given by (\ref{eqn:Konno1}), which
is the density function
first introduced by Konno to describe the limit
distributions of the two-component
one-dimensional quantum walks
in his weak limit-theorem \cite{Kon02,Kon05a}.
As a function of $y$, $|c^{(j)}_{m}(y)|^2$ is
separated into an even-function part and
an odd-function part.
For positive values of $m$, let
${\cal M}^{(j,m)}_{\rm even}(y/2m)=
\mbox{even-function part of} \, |c^{(j)}_{m}(y)|^2+|c^{(j)}_{-m}(y)|^2$
and
${\cal M}^{(j,m)}_{\rm odd}(y/2m)=
\mbox{odd-function part of} \, |c^{(j)}_{m}(y)|^2-|c^{(j)}_{-m}(y)|^2$.
Since $\mu(x; a)$ is an even function of $x$, 
(\ref{eqn:Ijm1}) gives
\begin{eqnarray}
&& I_{m}^{(j)}(2n)\nonumber\\
&=& \frac{1}{2m} \int_{-\infty}^{\infty}dy \,
y^{2n} \mu \left(\frac{y}{2m}; \cos \frac{\beta}{2} \right)
{\cal M}^{(j,m)}_{\rm even}\left(\frac{y}{2m}\right), \nonumber\\
&& I_{m}^{(j)}(2n-1) 
\nonumber\\
&=& \frac{1}{2m} \int_{-\infty}^{\infty}dy \,
y^{2n-1} \mu \left(\frac{y}{2m}; \cos \frac{\beta}{2} \right)
{\cal M}^{(j,m)}_{\rm odd}\left(\frac{y}{2m}\right)
\nonumber\\
\label{eqn:Ievenodd}
\end{eqnarray}
for $n=1,2,3, \cdots$.
Then (\ref{eqn:limit2}) implies that
\begin{eqnarray}
&& \lim_{t \to \infty} 
\Big\langle \Big( \frac{X^{(j)}_{t}}{t} \Big)^{r}
\Big\rangle \nonumber\\
&& \quad = \int_{-\infty}^{\infty} dy \, y^{r}
\sum_{m: 0 < m \leq j}
\frac{1}{2m} \mu \left( \frac{y}{2m}; \cos \frac{\beta}{2} \right)
\nonumber\\
&& \qquad \qquad \times
{\cal M}^{(j,m)}\left(\frac{y}{2m}\right)
\label{eqn:limit4}
\end{eqnarray}
for $r=1,2,3, \cdots$, where
\begin{equation}
{\cal M}^{(j,m)}(x)
={\cal M}^{(j,m)}_{\rm even}(x)+{\cal M}^{(j,m)}_{\rm odd}(x).
\label{eqn:calM}
\end{equation}

When $j$ is a positive integer,
({\it i.e.}, $(2j+1)$ is odd), 
the integral
\begin{equation}
J^{(j)}=\int_{-\infty}^{\infty} dy 
\sum_{m: 0 < m \leq j}
\frac{1}{2m} \mu \left( \frac{y}{2m}; \cos \frac{\beta}{2} \right)
{\cal M}^{(j,m)}\left(\frac{y}{2m}\right)
\label{eqn:Jj}
\end{equation}
is generally less than one,
since the contribution from the $m=0$ mode is
not included in the summation.
The difference
\begin{equation}
\Delta^{(j)}=1-J^{(j)}
\label{eqn:Delta0}
\end{equation}
gives the weight of a point mass at $y=0$ in the distribution.

The result is summarized as follows.
The long-time limit of
the pseudovelocity of $(2j+1)$-component
quantum walk is described by the probability measure,
which consists of the summation of
appropriately scaled Konno's density functions
with weight functions ${\cal M}^{(j,m)}(y/2m)$,
and a point mass at the origin 
with weight $\Delta^{(j)}$, if the number of
components $(2j+1)$ is odd,
that is 
\begin{equation}
\lim_{t \to \infty} 
\Big\langle \Big( \frac{X^{(j)}_{t}}{t} \Big)^{r}
\Big\rangle
= \int_{-\infty}^{\infty} dy \, y^{r}
\nu^{(j)}(y), \quad r=0,1,2, \cdots
\label{eqn:limit5}
\end{equation}
with
\begin{eqnarray}
\nu^{(j)}(y) &=& \sum_{m: 0 < m \leq j}
\frac{1}{2m} \mu \left( \frac{y}{2m}; \cos \frac{\beta}{2} \right)
{\cal M}^{(j,m)}\left(\frac{y}{2m}\right)
\nonumber\\
&&
+{\bf 1}_{\{(2j+1) \, \mbox{is odd}\}}
\Delta^{(j)} \delta(y).
\label{eqn:nu1}
\end{eqnarray}

\subsection{Polynomials ${\cal M}^{(j,m)}(x)$
representing parameter and initial-qudit dependence}

Using the formulas given in Appendix C and
the matrices $r^{(j)}$ in Appendix A, the weights
${\cal M}^{(j,m)}(x)$ and $\Delta^{(j)}$ 
in the limit distribution
(\ref{eqn:nu1}) are explicitly determined as follows
for $j=1/2, 1$ and 3/2, $0 < m \leq j$.
Set
\begin{equation}
\tau=\tan \frac{\beta}{2}.
\label{eqn:b}
\end{equation}
For a complex number $z$, ${\rm Re}\{z\}$ denotes
the real part of $z$.

\noindent \underline{$j=1/2$ case (two-component model)}
\begin{equation}
{\cal M}^{(1/2, 1/2)}(x)=1+{\cal M}^{(1/2, 1/2)}_{1} x, 
\label{eqn:M1/2}
\end{equation}
with
\begin{equation}
{\cal M}^{(1/2, 1/2)}_{1}=
-\{|q_{1/2}|^2-|q_{-1/2}|^2\}
+2 \tau \,
{\rm Re} \Big\{ q_{1/2} \overline{q}_{-1/2} e^{-\im \gamma} \Big\}.
\label{eqn:M1/2c1}
\end{equation}
When ${\cal M}^{(1/2, 1/2)}_{1}=0$
(resp. ${\cal M}^{(1/2, 1/2)}_{1} \not=0$),
the probability density function of 
limit distribution $\nu^{(1/2)}(y)$
is symmetric (resp. asymmetric)
\cite{Kon02,Kon05a,KFK05}.

\begin{widetext}
\noindent \underline{$j=1$ case (three-component model)}
\begin{equation}
{\cal M}^{(1, 1)}(x)={\cal M}^{(1,1)}_{0}
+{\cal M}^{(1, 1)}_{1} x
+{\cal M}^{(1,1)}_{2} x^2,
\label{eqn:M1}
\end{equation}
with
\begin{eqnarray}
{\cal M}^{(1,1)}_{0} &=& \frac{1}{2} 
\{ |q_{1}|^{2} +2|q_{0}|^{2} +|q_{-1}|^{2} \}
-{\rm Re} \Big\{ q_{1} \overline{q}_{-1} e^{-2\im \gamma} \Big\},
\nonumber\\
{\cal M}^{(1,1)}_{1} &=&  - \{ |q_{1}|^{2} -|q_{-1}|^{2} \} 
+\sqrt{2} \tau \,
{\rm Re} \Big\{ ( q_{1} \overline{q}_{0} 
+ q_{0} \overline{q}_{-1} ) e^{-\im \gamma} \Big\},
\nonumber\\
{\cal M}^{(1,1)}_{2} &=&   \frac{1}{2} 
 \{ |q_{1}|^{2} -2|q_{0}|^{2} +|q_{-1}|^{2} \} 
-\sqrt{2} \tau \,
 {\rm Re} \Big\{ ( q_{1} \overline{q}_{0} 
-q_{0} \overline{q}_{-1} ) e^{-\im \gamma} \Big\}
+(1+2\tau^2) \,
{\rm Re} \Big\{ q_{1} \overline{q}_{-1} e^{-2\im \gamma} \Big\},
\label{eqn:M1c1}
\end{eqnarray}
and
\begin{equation}
\Delta^{(1)}= 1 -\left\{
{\cal M}^{(1,1)}_{0} +\left(1-\sin \frac{\beta}{2} \right) 
{\cal M}^{(1,1)}_{2} \right\}.
\label{eqn:Delta1}
\end{equation}
The condition that the probability density function of 
limit distribution $\nu^{(1)}(y)$
is symmetric is given by ${\cal M}^{(1,1)}_{1}=0$.
Generally the point mass at the origin appears
with the weight $\Delta^{(1)}$ in the limit distribution.

\noindent \underline{$j=3/2$ case (four-component model)}
\begin{equation}
{\cal M}^{(3/2, 3/2)}(x)={\cal M}^{(3/2,3/2)}_{0}
+{\cal M}^{(3/2, 3/2)}_{1} x
+{\cal M}^{(3/2,3/2)}_{2} x^2
+{\cal M}^{(3/2,3/2)}_{3} x^3,
\label{eqn:M3/2A}
\end{equation}
and
\begin{equation}
{\cal M}^{(3/2, 1/2)}(x)={\cal M}^{(3/2,1/2)}_{0}
+{\cal M}^{(3/2, 1/2)}_{1} x +{\cal M}^{(3/2,1/2)}_{2} x^2
+{\cal M}^{(3/2,1/2)}_{3} x^3,
\label{eqn:M3/2B}
\end{equation}
with
\begin{eqnarray}
{\cal M}^{(3/2,3/2)}_{0} &=& \frac{1}{4} \Bigl\{ 
|q_{3/2}|^{2} +3|q_{1/2}|^{2} +3|q_{-1/2}|^{2} +|q_{-3/2}|^{2} \Bigr\}
-\frac{\sqrt{3}}{2} {\rm Re} 
\Bigl\{ (q_{3/2} \overline{q}_{-1/2} +q_{1/2} \overline{q}_{-3/2} ) 
e^{-2\im \gamma} \Bigr\}, \nonumber\\
{\cal M}^{(3/2,3/2)}_{1} &=&  -\frac{3}{4} \Bigl\{ 
|q_{3/2}|^{2} +|q_{1/2}|^{2} -|q_{-1/2}|^{2} -|q_{-3/2}|^{2} \Bigr\} 
-\frac{3}{2} \tau \,
{\rm Re} \Bigl\{ q_{3/2} \overline{q}_{-3/2} e^{-3\im \gamma}
-q_{1/2} \overline{q}_{-1/2} e^{-\im \gamma} \Bigr\} \nonumber\\
&& +\frac{\sqrt{3}}{2} \tau \,
{\rm Re} \Bigl\{ ( q_{3/2} \overline{q}_{1/2} 
+q_{-1/2} \overline{q}_{-3/2} ) e^{-\im \gamma} \Bigr\} 
 +\frac{\sqrt{3}}{2} 
{\rm Re} \Bigl\{ ( q_{3/2} \overline{q}_{-1/2} 
-q_{1/2} \overline{q}_{-3/2} ) e^{-2\im \gamma} \Bigr\},
\nonumber\\
{\cal M}^{(3/2,3/2)}_{2} &=& \frac{3}{4} \Bigl\{ 
|q_{3/2}|^{2} -|q_{1/2}|^{2} -|q_{-1/2}|^{2} +|q_{-3/2}|^{2} \Bigr\} 
-\sqrt{3} \tau \, {\rm Re} 
\Bigl\{ (q_{3/2} \overline{q}_{1/2} -q_{-1/2} \overline{q}_{-3/2} ) 
e^{-\im \gamma} \Bigr\} \nonumber\\
&& +\frac{\sqrt{3}}{2} (1+2\tau^2) \,
{\rm Re} 
\Bigl\{ (q_{3/2} \overline{q}_{-1/2} +q_{1/2} \overline{q}_{-3/2} ) 
e^{-2\im \gamma} \Bigr\},
\nonumber\\
{\cal M}^{(3/2,3/2)}_{3} &=& -\frac{1}{4} \Bigl\{ 
|q_{3/2}|^{2} -3|q_{1/2}|^{2} +3|q_{-1/2}|^{2} -|q_{-3/2}|^{2} \Bigr\} 
+\frac{1}{2} \tau (3+4\tau^2) \,
{\rm Re} \Bigl\{ q_{3/2} \overline{q}_{-3/2} e^{-3\im \gamma} \Bigr\} 
\nonumber\\
&& -\frac{3}{2} \tau \,
{\rm Re} \Bigl\{ q_{1/2} \overline{q}_{-1/2} e^{-\im \gamma} \Bigr\}
+\frac{\sqrt{3}}{2} \tau \,
{\rm Re} \Bigl\{ ( q_{3/2} \overline{q}_{1/2} 
+q_{-1/2} \overline{q}_{-3/2} ) e^{-\im \gamma} \Bigr\} \nonumber\\
&& -\frac{\sqrt{3}}{2} (1+2\tau^2) \,
{\rm Re} \Bigl\{ ( q_{3/2} \overline{q}_{-1/2} 
-q_{1/2} \overline{q}_{-3/2} ) e^{-2\im \gamma} \Bigr\},
\label{eqn:M3/2c1}
\end{eqnarray}
and with
\begin{eqnarray}
{\cal M}^{(3/2,1/2)}_{0} &=& \frac{1}{4} \Bigl\{ 
3|q_{3/2}|^{2} +|q_{1/2}|^{2} 
+|q_{-1/2}|^{2} +3|q_{-3/2}|^{2} \Bigr\} 
+\frac{\sqrt{3}}{2} {\rm Re} \Bigl\{ ( q_{3/2} \overline{q}_{-1/2} 
+q_{1/2} \overline{q}_{-3/2} ) e^{-2\im \gamma} \Bigr\}, \nonumber\\
{\cal M}^{(3/2,1/2)}_{1} &=&  -\frac{1}{4} \Bigl\{ 
3|q_{3/2}|^{2} -5|q_{1/2}|^{2} 
+5|q_{-1/2}|^{2} -3|q_{-3/2}|^{2} \Bigr\} 
+\frac{9}{2} \tau \, {\rm Re} \Bigl\{ 
q_{3/2} \overline{q}_{-3/2} e^{-3\im \gamma} \Bigr\} \nonumber\\
&& -\frac{1}{2} \tau \, {\rm Re} \Bigl\{ 
q_{1/2} \overline{q}_{-1/2} e^{-\im \gamma} \Bigr\} 
+\frac{\sqrt{3}}{2} \tau \, {\rm Re} \Bigl\{ 
( q_{3/2} \overline{q}_{1/2} +q_{-1/2} \overline{q}_{-3/2} )
e^{-\im \gamma} \Bigr\} \nonumber\\
&& -\frac{3 \sqrt{3}}{2} {\rm Re} \Bigl\{ 
( q_{3/2} \overline{q}_{-1/2} -q_{1/2} \overline{q}_{-3/2} )
e^{-2\im \gamma} \Bigr\}, \nonumber\\
{\cal M}^{(3/2,1/2)}_{2} &=&  -\frac{3}{4} \Bigl\{ 
|q_{3/2}|^{2} -|q_{1/2}|^{2} 
-|q_{-1/2}|^{2} +|q_{-3/2}|^{2} \Bigr\} 
+\sqrt{3} \tau \,
{\rm Re} \Bigl\{ ( q_{3/2} \overline{q}_{1/2} 
-q_{-1/2} \overline{q}_{-3/2} ) e^{-\im \gamma} \Bigr\}
\nonumber\\
&& -\frac{\sqrt{3}}{2} (1+2\tau^2) \,
{\rm Re} \Bigl\{ ( q_{3/2} \overline{q}_{-1/2} 
+q_{1/2} \overline{q}_{-3/2} ) e^{-2\im \gamma} \Bigr\}, \nonumber\\
{\cal M}^{(3/2,1/2)}_{3} &=& \frac{3}{4} \Bigl\{ 
|q_{3/2}|^{2} -3|q_{1/2}|^{2} 
+3|q_{-1/2}|^{2} -|q_{-3/2}|^{2} \Bigr\} 
-\frac{3}{2} \tau (3+4\tau^2) \,
{\rm Re} \Bigl\{ 
q_{3/2} \overline{q}_{-3/2} e^{-3\im \gamma} \Bigr\} 
\nonumber\\
&& +\frac{9}{2} \tau \, {\rm Re} \Bigl\{ 
q_{1/2} \overline{q}_{-1/2} e^{-\im \gamma} \Bigr\} 
-\frac{3 \sqrt{3}}{2} \tau \, {\rm Re} \Bigl\{ 
( q_{3/2} \overline{q}_{1/2} +q_{-1/2} \overline{q}_{-3/2} )
e^{-\im \gamma} \Bigr\} 
\nonumber\\
&& +\frac{3 \sqrt{3}}{2} (1+2\tau^2) \,
{\rm Re} \Bigl\{ 
( q_{3/2} \overline{q}_{-1/2} -q_{1/2} \overline{q}_{-3/2} )
e^{-2\im \gamma} \Bigr\}.
\label{eqn:M3/2c2}
\end{eqnarray}
If and only if 
${\cal M}^{(3/2, 3/2)}_{1}={\cal M}^{(3/2, 3/2)}_{3}=0$
and
${\cal M}^{(3/2, 1/2)}_{1}={\cal M}^{(3/2, 1/2)}_{3}=0$,
the probability density function of 
limit distribution $\nu^{(3/2)}(y)$ is 
symmetric.
\end{widetext}
These results imply that ${\cal M}^{(j,m)}(x)$ are
polynomials of $x$ of degree $2j$ and the
coefficients ${\cal M}^{(j,m)}_{k}, k=0,1, \cdots, 2j$
depend on $\beta$ and $\gamma$
through the functions $\tau=\tan(\beta/2)$
and $e^{-\im \gamma}$, but they do not on
$\alpha$.
It should be noted that their dependence
on initial qudit (\ref{eqn:phi01}) is complicated.
In other words, the limit distribution of pseudovelocity
of quantum walk is very sensitive to changes of initial qudit.

\section{COMPARISON WITH COMPUTER SIMULATIONS AND
CONCLUDING REMARKS}

In order to demonstrate the validity of the above results,
here we show comparison with computer simulation results.
In the following figures, 
Figs.\ref{fig:mkkFig2}-\ref{fig:mkkFig4},
the scattering thin lines indicate the
density of distribution of $X_{t}/t$ at time step $t=100$
obtained by computer simulation
and the thick lines the probability densities of 
limit distributions
$\nu^{(j)}(y)$ 
given in the previous section.
Note that if we integrate the density over
an interval $[a,b]$, then we obtain the
probability that the value of $X_t/t$ is 
in $[a,b]$.

\vskip 0.5cm
\noindent\underline{Two-component model} 

\begin{figure}[htpb]
\includegraphics[width=1.0\linewidth]{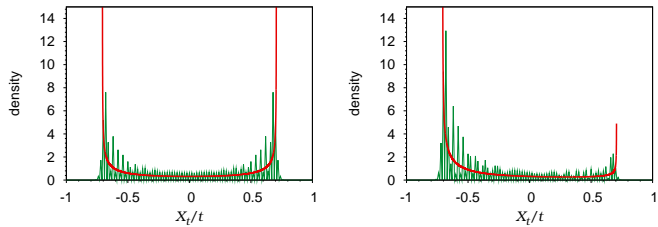}
\caption{(Color online)
Comparison between simulation results
and the probability densities of 
limit distributions for the two-component
model. (a) Symmetric and (b) asymmetric cases.
\label{fig:mkkFig2}}
\end{figure}

The result (\ref{eqn:M1/2}) with (\ref{eqn:M1/2c1})
is completely identified with the previous 
results \cite{Kon02,Kon05a,KFK05}.
Here we put 
$(\alpha, \beta, \gamma)=(0, -3 \pi/2, \pi)$.
Then we have
\begin{equation}
R^{(1/2)}=\frac{\im}{\sqrt{2}}
\left( \begin{array}{cc}
1 & 1 \cr 1 & -1 
\end{array} \right),
\label{eqn:exp1}
\end{equation}
which is the Hadamard matrix multiplied by $\im$.
(Remark that the factor $\im$ is irrelevant for limit distribution,
but by this factor $R^{(1/2)}$ is in SU(2),
see \cite{KFK05}.)
When we choose the initial qubit as
$^{t}\phi_{0}=(1+\im, 1-\im)/2$,
the limit distribution of pseudovelocity is symmetric
as shown by Fig.\ref{fig:mkkFig2} (a),
but when we choose
$^{t}\phi_{0}=(1+\im, 1+\im)/2$,
only changing the sign of imaginary part of
the second component, 
the limit distribution becomes asymmetric
as shown by Fig. \ref{fig:mkkFig2} (b).
When $(\alpha, \beta, \gamma)=(0, -3\pi/2, \pi)$,
the former initial qubit satisfies the condition
${\cal M}^{(1/2, 1/2)}_{1}=0$,
but the latter does not, where
${\cal M}^{(1/2, 1/2)}_{1}$ is given by
(\ref{eqn:M1/2c1}).
The shape of probability density function in 
limit distribution
is very sensitive to changes of initial qubit.

\vskip 0.5cm
\noindent\underline{Three-component model}

\begin{figure}[htpb]
\includegraphics[width=1.0\linewidth]{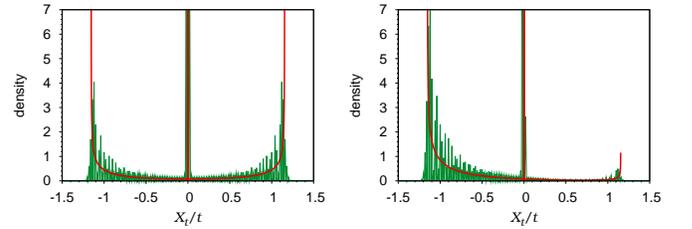}
\caption{(Color online)
Comparison between simulation results
and the probability densities of 
limit distributions for the three-component
model. (a) Symmetric and (b) asymmetric cases.
The perpendicular thick lines at the origin
indicate Dirac's delta functions.
\label{fig:mkkFig3}}
\end{figure}

If we set 
$(\alpha, \beta, \gamma)=(0, \arccos(-1/3), \pi)$,
the three-component quantum coin will be
\begin{equation}
R^{(1)}=\frac{1}{3}
\left( \begin{array}{ccc}
-1 & -2 & -2 \cr 
-2 & -1 & 2 \cr
-2 &  2 & -1 
\end{array} \right).
\label{eqn:exp2}
\end{equation}
Remark that it is similar to the Grover
matrix \cite{Gro97}, but it is not the same.
Figure \ref{fig:mkkFig3} shows the comparison
of simulation results and limit distributions
for (a) $^{t}\phi_{0}=(1-\im, 1+\im, 1-\im)/\sqrt{6}$,
which gives symmetric distribution,
and for (b) $^{t}\phi_{0}=(1-\im, 1-\im, 1-\im)/\sqrt{6}$,
which gives asymmetric distribution,
respectively.
It is readily checked that the case (a) satisfies the
condition ${\cal M}^{(1,1)}_{1}=0$ for
symmetric distribution.
In the three-component model,
Dirac's delta-function-type peak 
at the origin is usually observed in simulation.

\vskip 0.5cm
\noindent\underline{Four-component model}

\begin{figure}[htpb]
\includegraphics[width=1.0\linewidth]{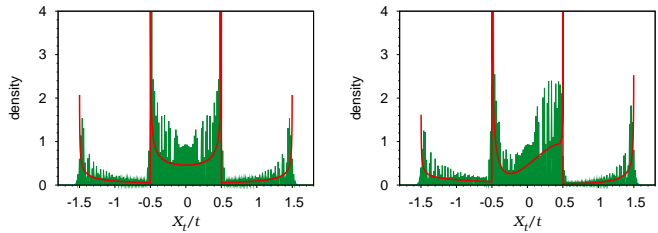}
\caption{(Color online)
Comparison between simulation results
and the probability densities of
limit distributions for the four-component
model. (a) Symmetric and (b) asymmetric cases.
\label{fig:mkkFig4}}
\end{figure}

We set
$(\alpha, \beta, \gamma)=(0, 2\pi/3, \pi)$
for the four-component model, which
corresponds to choosing the quantum coin as
\begin{equation}
R^{(3/2)}=\frac{\im}{8}
\left( \begin{array}{cccc}
1 & 3 & 3 \sqrt{3} & 3 \sqrt{3} \cr 
3 & 5 & \sqrt{3} & -3 \sqrt{3} \cr
3 \sqrt{3} & \sqrt{3} & -5 & 3 \cr
3 \sqrt{3} & -3 \sqrt{3} & 3 & -1
\end{array} \right).
\label{eqn:exp3}
\end{equation}
If we assume the initial qudit as
\begin{equation}
\phi_{0}=\frac{1}{2 \sqrt{5}} \left(
\begin{array}{c}
1+3 \im \cr
0 \cr
0 \cr
-3+\im 
\end{array} \right) \,
\mbox{and} \quad
\phi_{0}=\frac{1}{2 \sqrt{5}} \left(
\begin{array}{c}
1+3 \im \cr
0 \cr
0 \cr
-3-\im 
\end{array} \right),
\label{eqn:exp3b}
\end{equation}
the distributions are determined as
shown in Fig. \ref{fig:mkkFig4} (a) and
Fig. \ref{fig:mkkFig4} (b), respectively.

\vskip 0.5cm

We observe oscillatory behavior of density functions
of $X_{t}/t$ in computer simulations.
In general, as the time step $t$ increases,
the frequency of oscillation becomes higher.
But the convergence of any moments given by 
(\ref{eqn:limit5}) means that, if we smear out
the oscillatory behavior, the averaged lines
of density functions will be
well-described by those of the limit distributions
(\ref{eqn:nu1}),
which is the fact demonstrated by the above figures.

Now we discuss relations between our models and
other multi-component models and possible future problems.
Inui and Konno \cite{IK05} and
Inui {\it et al.} \cite{IKS05}
introduced a one-dimensional three-component quantum walk
and showed that a kind of localization phenomenon
occurs in their model by calculating
the long-time distribution of walker's position $X_{t}$.
They claimed that such a localization 
phenomenon results in a point mass at the origin, represented by
Dirac's delta function, in the limit distribution of $X_t/t$.
Though their model associated with the Grover matrix
does not belong to our family
of models, the structure of 
probability density function in 
limit distribution
obtained in our three-component model ($j=1$ case)
(\ref{eqn:nu1}) with $j=m=1$ and with (\ref{eqn:M1})
is very similar to the limit density
function of $X_{t}/t$, given by Eq.(16) in \cite{IKS05}.
Then the present work suggests that such a localization
phenomenon is universal for the models, in which
there is a positive probability for walker to stay at the
same position in each step.
Further study of localization phenomena
in quantum-walk models will be an interesting future problem.
The relation between the present models and
the tensor-product models of Brun {\it et al.} \cite{BCA03}
was already discussed in Sec.IV.
Similarity between the density functions
of walkers plotted in Fig.1 in their paper
and ours shown above will be explained by
reducibility of tensor-product models.
As demonstrated in Sec.IV. C, their $2^2$-dimensional
model will include our three-component model
and their $2^{3}$-dimensional model does our
four-component one, and thus
the density function shown by
Fig.1(b) (resp. Fig.1(c)) of Brun {\it et al.}
\cite{BCA03} has the same structure
with Fig.\ref{fig:mkkFig3}(b)
(resp. Fig.\ref{fig:mkkFig4}(b)) in the present paper.
Venegas-Andraca {\it et al.} \cite{VBBB05} also
reported quantum-walk models with entangle coins,
where a variety of distributions of walker's position
with multi-peak zones are plotted in figures.
They constructed quantum coins for multi-component
qudits by also considering tensor products
of the basic two-component quantum coins.
Systematic classifications of multi-component
quantum-walk models and patterns of their 
limit distributions will be an important future problem.
In the present paper, we gave the explicit expressions
of ${\cal M}^{(j,m)}(x)$ for $j=1/2, 1$ and $3/2$.
Here we gave only key formulas in Appendix C
for calculation of them, but the present result
implies that ${\cal M}^{(j,m)}(x)$ are generally
given by polynomials.
We hope that further study of ${\cal M}^{(j,m)}$
is a promising subject in the field of quantum walks.
As reported in \cite{GJS04,MBSS02,Ken06,IKK04,GJS05,VBBB05},
multi-component models have been studied 
to simulate quantum walks on the plane,
in the higher-dimensional lattices $\Z^{d}$, or 
on general graphs.
The present work suggests that
the group-theoretical investigation will be useful
to perform systematic study of
such extended models of quantum walks
and quantum processes.

\begin{acknowledgments}
TM and MK thank S. Fujino for his collaboration
of the present work at the beginning.
MK acknowledges useful comments on the manuscript
by N. Inui.
MK also thanks M. Sato and K. Watabe
for preparing figures in the final version
of the manuscript.
This work was partially supported by
the Grant-in-Aid for Scientific Research
(KIBAN-C, No. 17540363) of Japan Society 
for the Promotion of Science.
\end{acknowledgments}

\begin{widetext}
\appendix
\section{The matrices $r^{(j)}(\beta)$ for $j=1/2, 1, 3/2$}
The explicit expressions of $r^{(j)}(\beta)$
are readily derived from the Wigner formula (\ref{eqn:rj1})
as follows for $j=1/2, 1, 3/2$.
Let
$
c=\cos (\beta/2), s=\sin(\beta/2).
$
Then
\begin{eqnarray}
\label{eqn:r1/2}
r^{(1/2)}(\beta)
&=& \left( \begin{array}{cc}
c & - s \cr
s & c 
\end{array} \right), \\
\label{eqn:r1}
r^{(1)}(\beta)
&=& \left( \begin{array}{ccc}
c^2 & - \sqrt{2} c s & s^2 \cr
\sqrt{2} c s &  2c^2-1 & -\sqrt{2} c s \cr
s^2 & \sqrt{2} c s &c^2 
\end{array} \right), \\
\label{eqn:r3/2}
r^{(3/2)}(\beta) \nonumber\\
&=&
\left( \begin{array}{llll}
c^3 & -\sqrt{3}c^2 s  & \sqrt{3} c s^2 & - s^3 \cr
\sqrt{3} c^2 s & -2 c s^2 +c^3 & s^3 -2 c^2 s
& \sqrt{3} c s^2 \cr
\sqrt{3} c s^2 & -s^3 +2 c^2s & -2 c s^2 +c^3
& -\sqrt{3} c^2 s \cr
s^3 & \sqrt{3} c s^2 & \sqrt{3} c^2 s & c^3
\end{array} \right).
\nonumber\\
\end{eqnarray}

\section{PROOF OF EQUATIONS (\ref{eqn:decom1}) WITH (\ref{eqn:decom2})}
By (\ref{eqn:Vjk1}), 
it is enough to prove the equality
\begin{equation}
R^{(j)}(\alpha, \beta, \gamma)
=R^{(j)}(\phi, \theta, 0)
R^{(j)}(-p, 0, 0)
[R^{(j)}(\phi, \theta,0)]^{\dagger}
\label{eqn:eqB1}
\end{equation}
with relations
\begin{equation}
\frac{1}{2}(\alpha-\gamma)=\phi+\frac{\pi}{2}, \quad
\tan \frac{1}{2}(\alpha+\gamma)=-\tan\frac{p}{2} \cos \theta, \quad
\sin \frac{\beta}{2}=\sin \frac{p}{2} \sin \theta.
\label{eqn:eqB2}
\end{equation}
The equality (\ref{eqn:eqB1}) is a matrix representation 
of the equality which will hold among the rotations
\begin{equation}
\hat{R}(\alpha, \beta, \gamma)
=\hat{R}(\phi, \theta, 0)
\hat{R}(-p, 0, 0)
[\hat{R}(\phi, \theta,0)]^{\dagger}.
\label{eqn:eqB3}
\end{equation}
Since rotation matrices defined by (\ref{eqn:Rj1})
with $j=1/2, 1, 3/2, \cdots$
are faithful representations of rotations,
it may be enough to prove the equality (\ref{eqn:eqB1})
for the simplest case, $j=1/2$, since it implies
(\ref{eqn:eqB3}).
By direct calculation, we have
\begin{eqnarray}
&& R^{(1/2)}(\phi, \theta, 0)
R^{(1/2)}(-p, 0, 0)
[R^{(1/2)}(\phi, \theta, 0)]^{\dagger}
\nonumber\\
&=& \left( \begin{array}{cc}
e^{-\im \phi/2} \cos(\theta/2) & - e^{-\im \phi/2} \sin(\theta/2) \cr
e^{\im \phi/2} \sin(\theta/2) & e^{\im \phi/2} \cos(\theta/2) 
\end{array} \right) 
\left( \begin{array}{cc}
e^{\im p/2} & 0 \cr 0 & e^{-\im p/2} \end{array} \right)
\left( \begin{array}{cc}
e^{\im \phi/2} \cos(\theta/2) & e^{-\im \phi/2} \sin(\theta/2) \cr
-e^{\im \phi/2} \sin(\theta/2) & e^{-\im \phi/2} \cos(\theta/2) 
\end{array} \right)
\nonumber\\
&=& \left( \begin{array}{cc}
\cos(p/2)+\im \sin(p/2)\cos \theta &
\im \sin(p/2) \sin \theta e^{-\im \phi} \cr
\im \sin(p/2) \sin \theta e^{\im \phi} &
\cos(p/2)-\im \sin(p/2) \cos \theta 
\end{array} \right).
\nonumber\\
\label{eqn:eqB4}
\end{eqnarray}
Comparing it with
$$
R^{(1/2)}(\alpha, \beta, \gamma)
=\left( \begin{array}{cc}
e^{-\im(\alpha+\gamma)/2} \cos(\beta/2) &
-e^{-\im(\alpha-\gamma)/2} \sin(\beta/2) \cr
e^{\im(\alpha-\gamma)/2} \sin(\beta/2) &
e^{\im(\alpha+\gamma)/2} \cos(\beta/2)
\end{array} \right),
$$
we have the equations
\begin{eqnarray}
&& \cos \frac{p}{2}=\cos \frac{\beta}{2} 
\cos \frac{1}{2}(\alpha+\gamma),
\quad
\sin \frac{p}{2} \cos \theta = - \cos \frac{\beta}{2} 
\sin \frac{1}{2}(\alpha+\gamma), \nonumber\\
&& \sin \frac{p}{2} \sin \theta \sin \phi
=-\sin \frac{\beta}{2} \cos \frac{1}{2}(\alpha-\gamma),
\quad 
\sin \frac{p}{2} \sin \theta \cos \phi
= \sin \frac{\beta}{2} \sin \frac{1}{2}(\alpha-\gamma),
\label{eqn:eqB5}
\end{eqnarray}
from which (\ref{eqn:eqB2}) is derived.

\section{FORMULAS}
Inserting (\ref{eqn:Rj2}) with (\ref{eqn:rj1}) 
into (\ref{eqn:Cj1})
and taking square of the complex variable, 
we have the expression
\begin{eqnarray}
|C_{m}^{(j)}(k)|^2
&=& \sum_{m_1=-j}^{j} \sum_{m_2=-j}^{j} q_{m_1} 
\overline{q}_{m_2} e^{\im (m_1-m_2) \phi(k)}
\sum_{\ell_1} \sum_{\ell_2} 
\Gamma(j, m_1, m, \ell_1)
\Gamma(j, m_2, m, \ell_2) \nonumber\\
&& \qquad \times \left( \cos \frac{\theta(k)}{2} 
\right)^{4j+m_1+m_2-2m-2\ell_1-2\ell_2}
\left( \sin \frac{\theta(k)}{2} 
\right)^{2 \ell_1+2\ell_2+2m-m_1-m_2}.
\label{eqn:Csq2}
\end{eqnarray}
From (\ref{eqn:mapA3})-(\ref{eqn:mapA5}), 
the following relations are derived,
\begin{eqnarray}
&& \cos \theta(k)= - \cos \frac{\beta}{2} \sin \chi,
\nonumber\\
&& \sin \theta(k) e^{\im \phi(k)}=
\left( \sin \frac{\beta}{2} \sin \chi
- \im \cos \chi \right) e^{- \im \gamma}.
\label{eqn:Relation1}
\end{eqnarray}
Then, through the change of variable (\ref{eqn:y1}),
we have
\begin{eqnarray}
\cos^2 \frac{\theta(k)}{2}
&=& \frac{1}{2} \left( 1-\frac{y}{2m} \right),
\nonumber\\
\sin^2 \frac{\theta(k)}{2}
&=& \frac{1}{2} \left( 1+ \frac{y}{2m} \right),
\nonumber\\
\sin \frac{\theta(k)}{2} \cos \frac{\theta(k)}{2}
e^{\im \phi(k)}
&=& \frac{1}{2} \left\{
\frac{y}{2m} \tan \frac{\beta}{2}
- \im \sqrt{1-\left(\frac{y}{2m}\right)^2
\sec^2 \frac{\beta}{2}} \right\} e^{-\im \gamma}.
\label{eqn:Relation2}
\end{eqnarray}
By (\ref{eqn:Relation2}) we can perform the
transformations $|C^{(j)}_{m}(k)|^2
\mapsto |\hat{C}^{(j)}_{m}(\chi)|^2
\mapsto |c^{(j)}_{m}(y)|^2$,
and ${\cal M}^{(j,m)}(y/2m)$'s are obtained.

In order to calculate (\ref{eqn:Delta1}),
we have used the following integral formulas,
\begin{eqnarray}
\int_{0}^{2 \cos(\beta/2)}
\frac{1}{\sqrt{\cos^2(\beta/2)-(y/2)^2}} dy &=& \pi,
\nonumber\\
 \int_{0}^{2 \cos(\beta/2)}
\frac{1}{\{1-(y/2)^2\} \sqrt{\cos^2(\beta/2)-(y/2)^2}} dy 
&=& \frac{\pi}{\sin(\beta/2)}.
\label{eqn:intform}
\end{eqnarray}
\end{widetext}


\end{document}